\documentstyle[12pt]{article}
\def\P{{\rm P}}
\def\QG{{\rm QG}}

\begin{document}
\small

\begin{flushright}
ACT-10/98 \\
CTP-TAMU-40/98 \\
NEIP-98-014\\
astro-ph/9810483 \\ 
\end{flushright}

\begin{centering}
\bigskip
{\large {\bf Sensitivity of Astrophysical Observations to
             Gravity-Induced Wave Dispersion in Vacuo}} \\
\bigskip
{\bf G. Amelino-Camelia$^{a}$},
{\bf John Ellis$^{b}$},
{\bf N.E. Mavromatos$^{c}$}, \\
{\bf D.V. Nanopoulos}$^{d}$ and 
{\bf Subir Sarkar}$^{c}$ \\
\bigskip \noindent 
$^a$ Institut de Physique, Universit\'e de Neuch\^atel,
     CH-2000 Neuch\^atel, Switzerland \\
$^b$ Theory Division, CERN, CH-1211, Geneva, Switzerland, \\
$^c$ Theoretical Physics, University of Oxford, 1 Keble Road, Oxford
     OX1 3NP, UK \\
$^d$ Dept. of Physics, Texas A \& M University, College Station, 
     TX 77843-4242, and \\
     HARC, The Mitchell Campus, Woodlands, TX 77381, USA; \\
     Academy of Athens, 28 Panepistimiou Avenue, Athens 10679, Greece. \\
\bigskip 
{\bf Abstract} \\
\bigskip
\end{centering}
{\bf We discuss possible signatures of quantum gravity for the
propagation of light, including an energy-dependent velocity
(refractive index), dispersion in velocity at a given energy, and
birefringence. We also compare the sensitivities of different
astrophysical observations, including BATSE data on GRB~920229,
BeppoSAX data on GRB~980425, the possible HEGRA observation of
GRB~920925c, and Whipple observations of the active galaxy
Mrk~421. Finally, we discuss the prospective sensitivities of AMS and
GLAST.}
\bigskip

We pointed out recently~\cite{AEMNS} that observations of $\gamma$-ray
bursters (GRBs) can be used to test models of quantum gravity in which
the velocity of light depends on the photon energy $E$. This is
because GRBs combine large distances, $z\sim\,1$, with fine time
structure, $\Delta\,t\sim{\cal O}(10^{-3})$~s. Assuming a linear
dependence of the velocity, $v=c(1 \pm(E/E_{\QG})$, we suggested that
effective quantum gravity scales $E_{\QG}$ up to the Planck scale,
$M_\P\sim10^{19}$~GeV, can in principle be probed using GRBs, and
argued that extant data already has a sensitivity to
$E_{\QG}\sim10^{16}$~GeV. Subsequently there have been several
interesting developments which we comment on here.

\medskip
The energy-dependent dispersion effect~\cite{AEMNS} can be interpreted
as a refractive index {\it in vacuo} induced by quantum
gravity. Another possibility is a statistical velocity spread among
photons of the same energy $E$; higher-order quantum-gravity effects
may induce $\delta\,v\,(E)\propto\,E$ which can be probed in a similar
manner. Arguments have also been given in the context of a loop
representation of canonical quantum gravity~\cite{GP} for the
possibility of energy-dependent birefringence:
$v_{\pm}=c(1\pm\,E/E_{\QG})$ for $\pm$ photon helicity states, as well
as a linear energy dependence of the refractive index {\it in
vacuo}. We note that birefringence effects are, in general,
characteristic of theories with (spontaneous) violation of Lorentz
invariance~\cite{kostel}; probing such effects might be possible with
polarization measurements of GRBs. In each case, the figure of merit
for comparing the sensitivities of different sources (at distance $D$)
is the parameter $\eta\equiv(DE/cM_P)/\Delta\,t$, which we now compare
for various observations.

\medskip
We had previously discussed GRB~920229, which exhibited
micro-structure in its burst at energies up to $\sim200$~keV. We
estimated conservatively that a detailed time-series analysis might
reveal coincidences in different BATSE energy bands on a time-scale
$\sim10^{-2}$~s, which would yield $\eta\sim10^{-3}$ i.e. sensitivity
to $E_{\QG}\sim10^{16}$~GeV, assuming that GRB~920229 has an
(apparently) typical redshift of ${\cal O}(1)$, corresponding to a
distance of $\sim3000$~Mpc. A similar sensitivity might be obtainable
with GRB~980425, given its likely identification with the unusual
supernova 1998bw \cite{sngrb}. This is known to have taken place at a
redshift $z=0.0083$ corresponding to a distance $D\sim40$~Mpc (for a
Hubble constant of 65 km\,sec$^{-1}$Mpc$^{-1}$) which is rather
smaller than a typical GRB distance. However GRB~980425 was observed
by BeppoSAX at energies up to $1.8$~MeV, which gains back an order of
magnitude in the overall figure of merit. If a time-series analysis
were to reveal structure at the $\Delta\,t\sim10^{-3}$~s level, which
is typical of many GRBs~\cite{variation}, it would yield the same
sensitivity as GRB~920229, with the advantage that its redshift is
known rather than estimated.

\medskip
A more speculative possibility is offered by GRB~920925c, observed by
WATCH~\cite{watch} and possibly in high-energy $\gamma$ rays by the
HEGRA/AIROBICC array above $20$~TeV~\cite{hegra}. Several caveats are
in order: taking into account the appropriate trial factor, the
confidence level for the signal seen by HEGRA to be related to
GRB~920925c is only $99.7\%$ ($\sim2.7\sigma$), the reported
directions differ by 9$^0$, and the redshift of the source is
unknown. Nevertheless, the potential sensitivity is impressive. The
events reported by HEGRA range up to $ E\sim200$~TeV, and the
correlation with GRB~920925c is within $\Delta\,t\sim200$~s. Making
the reasonably conservative assumption that GRB~920925c occurred no
closer than GRB~980425, viz. $\sim40$~Mpc, we find a minimum figure of
merit $\eta\sim\,1$, corresponding to a possible sensitivity to
$E_{\QG}\sim10^{19}$~GeV, modulo the caveats listed above. Even more
spectacularly, several of the HEGRA GRB~920925c candidate events
occurred within $\Delta\,t\sim1$~s, providing a potential sensitivity
even two orders of magnitude higher.

\medskip
Other astrophysical objects may also provide sensitive experimental
probes of the quantum-gravity effects discussed
earlier~\cite{AEMNS}. In particular, a strong limit has been
extracted~\cite{Whipple} using data from the Whipple telescope on a
TeV $\gamma$-ray flare associated with the active galaxy Mrk~421.
This object has a redshift of 0.03 corresponding to a distance of
$\sim100$~Mpc. Four events with $\gamma$-ray energies above 2~TeV have
been observed within a period of $280$~s. These provide a figure of
merit somewhat larger than we had previously suggested for GRB~920229,
and with the capital advantage that the redshift of the source is {\it
known}. Thus a definite limit $E_{\QG}>4\times10^{16}$~GeV was
derived~\cite{Whipple}.

\medskip
What of the future? A new generation of orbiting spectrometers, e.g.,
AMS~\cite{AMS} and GLAST~\cite{GLAST}, are being developed, whose
potential sensitivities are very impressive. For example, assuming a
$E^{-2}$ energy spectrum, GLAST would expect to observe about 25 GRBs
per year at photon energies exceeding 100~GeV, with time resolution of
microseconds. AMS would observe a similar number at $E>10$~GeV
with time resolution below 100~nanoseconds. For a nominal redshift of
$z\sim0.1$ corresponding to a distance of $D\sim300$~Mpc, the expected
time-delay at 10-100 GeV would be $\Delta\,t\sim30-300$~ms for
$E_{\QG}\sim10^{19}$~GeV, which should be simple to detect. We
conclude that these missions would be adequate to exclude or establish
the existence of quantum-gravity effects such as a refractive index,
statistical dispersion or birefringence, if they appear with a linear
dependence scaled by $E/M_{\P}$. We believe this adds significantly to
the scientific case for these experiments since they should be able to
constrain significantly interesting theories of fundamental physics.

\vfill \noindent {\bf Acknowledgements}

We thank Steven Biller, Elliot Bloom, Hans Hofer, Luis Padilla and
Sergei Sazonov for useful communications. This work was supported in
part by a grant from the Swiss National Science Foundation (G.A.-C.),
a P.P.A.R.C. advanced fellowship (N.E.M.), and D.O.E. Grant
DE-FG03-95-ER-40917 (D.V.N.).

\newpage
\baselineskip 12pt plus .5pt minus .5pt
 

\begin{thebibliography}{99}

\bibitem{AEMNS}
 G. Amelino-Camelia, J. Ellis, N.E. Mavromatos, D.V. Nanopoulos and 
  S. Sarkar, Nature 393, 763-765 (1998).

\bibitem{GP} 
 R. Gambini and G. Pullin, eprint gr-qc/9809038.

\bibitem{kostel} 
 D. Colladay and V. A. Kosteleck\'y , eprint hep-ph/9809521. 

\bibitem{Whipple} 
 S.D. Biller {\it et al.}, eprint gr-qc/9810044.

\bibitem{sngrb} 
 S.R. Kulkarni {\em et al}, Nature 395, 663-669 (1998);
 T.J. Galama {\em et al}, Nature 395, 670-672 (1998).

\bibitem{variation} 
 K.C. Walker, B.E. Schaefer and E.E. Fenimore, eprint astro-ph/9810271.

\bibitem{watch} 
 S.Y Sazonov {\em et al}, Astron. Astrophys. Suppl. 129, 1-8 (1998).

\bibitem{hegra} 
 L. Padilla {\em et al}, Astron. Astrophys. 337, 43-50 (1998). 

\bibitem{AMS} 
 AMS Collaboration, S. Ahlen {\em et al}, 
  Nucl. Instrum. Meth. A350, 351-367 (1994).

\bibitem{GLAST} 
 GLAST Team, E.D. Bloom {\em et al},
  {\it Proc. Intern. Heidelberg Workshop on TeV Gamma-ray
  Astrophysics}, eds. H.J. Volk and F.A. Aharonian (Kluwer, 1996)
  pp.109--125.

\end{thebibliography}
\end{document}